\documentclass[12pt,preprint]{aastex}
\usepackage{txfonts}
\usepackage[T1]{fontenc}
\usepackage{graphicx}
\usepackage{enumerate}
\usepackage{appendix} 

\usepackage{amssymb}
\usepackage{multirow}
\usepackage{rotating}
\usepackage{ctable}
\usepackage{morefloats}
\usepackage{longtable}
\usepackage{bigstrut}
\usepackage[colorlinks, citecolor=blue]{hyperref}
\usepackage{graphicx,epsfig,fancyhdr,epsf,txfonts,natbib,epstopdf}
\usepackage{latexsym,graphicx,bbm}
\usepackage{float}
\usepackage{natbib}
\usepackage{booktabs}
\usepackage{ae,aecompl}

\begin{document}
\title{WES - Weihai Echelle Spectrograph}

\author{Dong-Yang Gao\altaffilmark{1}, Hang-Xin Ji\altaffilmark{2}, Chen Cao\altaffilmark{1}, Shao-Ming Hu\altaffilmark{1}, Robert A. Wittenmyer\altaffilmark{3,4}, Zhong-Wen Hu\altaffilmark{2}, Frank Grupp\altaffilmark{5}, Hanna Kellermann\altaffilmark{5}, Kai Li\altaffilmark{1}, Di-Fu Guo\altaffilmark{1} }

\altaffiltext{1}{Shandong Provincial Key Laboratory of Optical Astronomy 
and Solar-Terrestrial Environment, Institute of Space Sciences, Shandong 
University,Weihai, 264209, China (e-mail: gaodongyang@sdu.edu.cn (Dong-Yang Gao))}
\altaffiltext{2}{Nanjing Institute of Astronomical Optics \& Technology ,National Astronomical Observatories ,Chinese Academy of Sciences, Nanjing 210042, China}
\altaffiltext{3}{School of Physics and Australian Centre for 
Astrobiology, University of New South Wales, Sydney 2052, Australia}
\altaffiltext{4}{Computational Engineering and Science Research Centre, 
University of Southern Queensland, Toowoomba, Queensland 4350, 
Australia}
\altaffiltext{5}{Universit\"ats Sternwarte M\"unchen, Scheinerstr. 1, 81679 M\"unchen, Germany}

\begin{abstract}

The Weihai Echelle Spectrograph (WES) is the first fiber-fed echelle 
spectrograph for astronomical observation in China. It is primarily used 
for chemical abundance and asteroseismology studies of nearby bright 
stars, as well as radial velocity detections for exoplanets. The optical 
design of WES is based on the widely demonstrated and well-established 
white-pupil concept. We describe the WES in detail and present some 
examples of its performance. A single exposure echelle image covers the 
spectral region 371-1,100 nm in 107 spectral orders over the rectangular 
CCD. The spectral resolution $R=\lambda/\Delta\lambda$ changes from 40,600 to 57,000 through adjusting the entrance slit width from full to 2.2 
pixels sampling at the fiber-exit. The limiting magnitude scales to 
$V=8$ with a signal-to-noise ratio (SNR) of more than 100 in $V$ for an hour exposure, at the spectral resolution R$\approx$40,000 in the median 
seeing of 1.7$^{\prime\prime}$ at Weihai Observatory (WHO) for the 
1-meter telescope. The radial velocity (RV) measurement accuracy of WES 
is estimated to be $<$10 m/s in 10 months (302 days) and better than 15 
m/s in 4.4 years (1,617 days) in the recent data processing.

\end{abstract}


\section{Introduction}

Weihai Observatory (WHO) of Shandong University is located at 
$122^\circ02^\prime58.6^{\prime\prime} E, 37^\circ32^\prime09.3^{\prime\prime} N$.  Made 
by the German APM Company, the WHO 1-meter telescope (WHOT), with an f/8 
classic Cassgerain design and equatorial fork mount mechanical 
structure, was installed in 2007 June. Standard Johnson-Cousins UBVRI 
and Stromgren \textit{uvby} filters are mounted at the Cassegrain focal 
plane for the photometric system. On more than 85\% of observational 
nights, the seeing is better than 2.0$^{\prime\prime}$ and the median 
seeing is 1.7$^{\prime\prime}$ at WHO (Hu et al. 2014)\footnote{The 
values of seeing were measured by the full width at half maximum (FWHM) 
of sources in the images during photometric observations.}. Sky 
brightness is influenced primarily by urban light pollution, which 
constrains the limiting magnitude of the photometry system. The faintest 
sky brightness is about 19.0 mag arcsec$^{−2}$ and the limiting 
magnitude with SNR of 100 and exposure of 300 s is 16.2 for V band at 
WHO (Guo et al. 2014; Hu et al. 2014). The flexible scheduling character 
of WHOT allows for long-term monitoring programs with a focus on stellar 
astrophysics that mainly call for high-quality and high-resolution 
spectra. And the high resolution spectroscopic observations are less 
influenced than photometric observations because of high optical 
dispersion.

Since the discovery of 51 Peg b, the first exoplanet orbiting a 
solar-like star (Mayor \& Queloz 1995), well over 2 900 exoplanets have 
been confirmed\footnote{http://exoplanets.org}.  The detection and 
characterisation of exoplanets are among the most popular topics in 
modern astronomy and astrophysics.  The common methods for searching for 
exoplanets are radial velocity (RV) detection, transit detection and 
gravitational microlensing.  High dispersion optical spectroscopic 
observations and high precision radial velocity (PRV) measurements on 
exoplanets are important for searching for new planetary systems around 
dwarf and giant stars, and for determining the architecture of 
exoplanetary systems (Sato et al. 2005a). Many groups around the world 
have built successful spectrographs for these purposes, such as ELODIE 
(Baranne et al. 1996) with which the first exoplanet (51 Peg b) was 
discovered, FOCES (Pfeiffer et al. 1998), FEROS (Kaufer et al. 1998), 
CORALIE (Queloz \& Mayor 2001), HARPS (Pepe et al. 2000), HERMES (Raskin 
et al. 2011) and CAF\'{E} (Aceituno et al. 2013).

As a complement to the current photometric measurement, the WES, 
designed and built by Nanjing Institute of Astronomical Optics \& 
Technology (NIAOT), was installed in 2010 to exploit the telescope 
infrastructure. The design of WES was guided by the scientific need for 
a high-resolution spectrograph with a high optical efficiency and very 
high wavelength stability. Like many high resolution spectrographs, the 
optical layout of WES is based on echelle grating, white pupil and 
fiber-fed design. The white pupil design with smaller cross-disperser 
size can reduce the scattered light emanating from the echelle gratings 
and avoids vignetting (Harrison et al. 1976; Baranne 1988; Pfeiffer et 
al. 1998). The optical fibers can transmit the light of celestial object 
to a bench-mounted spectrograph installed in a stable environment which 
permits higher radial velocity measurement precison. The spectroscopic 
observation can provide data for the studies of metal abundance of 
stars, exoplanet mass determination, asteroseismology, etc.

We give a description of the WES in Sect. 2, including the telescope 
interface module, fiber optics, optical layout, and operating 
environment. In Sect. 3, we describe the observations and data 
reduction. In Sect. 4 we present some data showing the performance of 
WES. The final section carries a brief summary and future prospects.

\section{Instrument}

The WES is designed for studies of metal abundance, stellar activity, 
and asteroseismology on nearby bright stars. In particular, it is 
specialized so as to have high RV measurement precision to search for 
and characterise giant exoplanets orbiting bright stars. Precise radial 
velocity measurements are needed to detect the small amplitude reflex 
motion of an exoplanet's host star.  So the most important design goal of 
WES is radial velocity measurement accuracy and stability (e.g., the 
stability in short-period is $\sim$10 m/s for a 6th magnitude star with 
the exposure of 10 min and the measurement error of 2 m/s.). Similar to 
most high resolution spectrographs, WES is also a cross-dispersed 
echelle spectrograph with a large 2K$\times$2K CCD chip which can image 
the complete visible spectrum in one exposure. WES adopts a fiber-fed 
mode and is installed in a very stable environment. The environment and 
mechanical design focus on holding the temperature constant and avoiding 
mechanical vibrations, which are critical for the high accuracy and 
stability of wavelength calibration. An iodine (I${_2}$) cell can be 
placed into the beam to determine the instrumental profile (IP), as is 
commonly used in exoplanet research (Butler et al. 1996; Valenti et al. 
1995).

\subsection{Telescope interface module}

\begin{figure}
\begin{center}
\includegraphics[angle=0,scale=0.3]{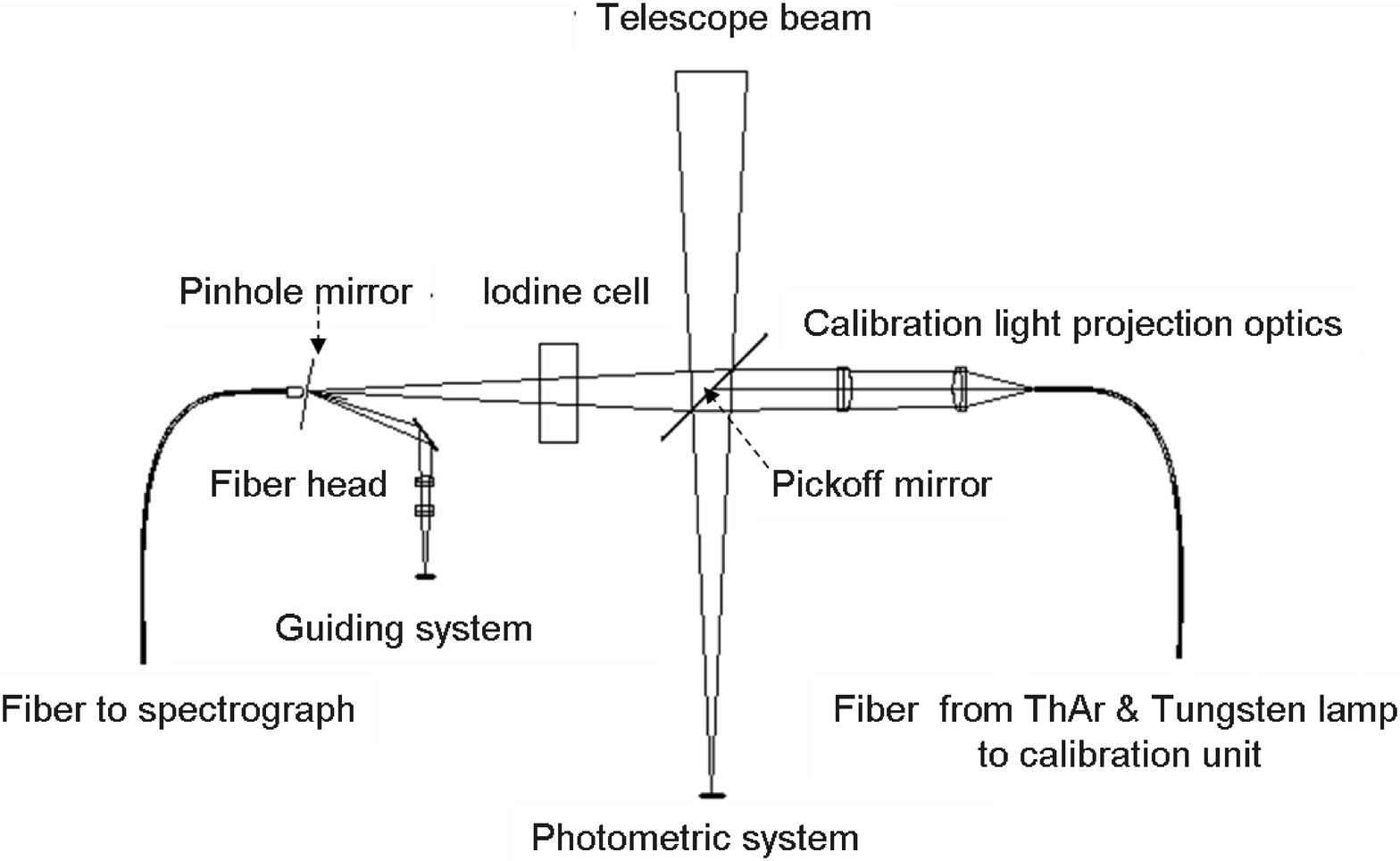}
\caption{A schematic diagram of interface to the telescope.}
\label{FigInterface}
\end{center}
\end{figure}

The telescope interface module mounted at the Cassegrain focal plane has 
two functions, one is for auto star guiding and injecting the light of 
flat field (Tungsten) and Thorium-Argon (ThAr) calibration lamp into the 
fiber-head, the other is to change the observation mode between 
photometric system (Hu et al. 2014) and spectroscopic system. 
Fig.~\ref{FigInterface} shows the interface to telescope including the 
fiber-head, the telescope guiding system and the calibration unit.

A 45$^\circ$ tilt mirror mounted on a compact motorized translation 
stage can be moved into and out of the optical light path in front of 
the telescope focal plane, at the center of the telescope interface. The 
light from telescope is guided to the fiber-head when the 45$^\circ$ 
pickoff mirror is moved in the optical light path. When the pickoff 
mirror is moved out, the light from telescope goes into the photometric 
system. The light of ThAr or flat-field lamp from the calibration unit 
simulates the focal ratio of telescope (f/8) by the projection optics. A 
black circular shading is attached on a lens in the projection optics 
for simulating pupil of the telescope. The simulation of the light 
between telescope and calibration unit is very important for PRV 
measurement because of instrument profile stability (Valenti et al. 
1995).

The fiber-head includes the entrance pinhole mirror mounted at the focal 
plane of telescope and the input microlens glued on the top of fiber. 
The microlens is used to re-image the telescope pupil onto the fiber 
which will be described in Sect 2.2. The diameter of the pinhole is 
decided by the median seeing at WHO which is 1.7$^{\prime\prime}$ (Hu et 
al. 2014). Too much light will be lost if the diameter is too small. If 
the diameter of the pinhole is too large, all of the star image will 
fall into the pinhole, then it is not feasible to monitor the star 
position and we cannot make sure that the illumination of fiber-head is 
uniform. The diameter of the pinhole is 0.1 mm corresponding to 
2.6$^{\prime\prime}$ in the sky. The throughput of the pinhole is 80\% 
at seeing=1.7$^{\prime\prime}$ condition calculated using the method of 
Pfeiffer et al. (1998).

The pinhole mirror is slightly inclined to reflect the star guiding 
image of $6^\prime\times6^\prime$ field of view at the focal plane via 
another inclined mirror through guiding optics to a SBIG ST2000 XMI CCD. 
The guiding optics is telecentric with characteristics of constant 
magnification and geometry in all field of view, which is good for 
target identification and automatic guiding.  Fig.~\ref{FigFiber}b shows 
an image of the guiding field.  The guiding can be implemented on the 
overflow annulus of the science target when only the target appears in 
the field of view. We can also use another star in the field of view for 
manual or automatic offset guiding in the situation as shown in 
Fig.~\ref{FigFiber}b. The telescope guiding also affects the 
instrumental profile (IP) of WES. A software for automatic guiding has 
been developed. Because of the good tracking accuracy which can reach 
0.6$^{\prime\prime}$ (RMS) in 10 min of blind guiding (Hu et al. 2014), 
the position of the target star is almost always centered on the pinhole 
during the long exposure ($\sim$30-60 minutes).

The iodine cell can be moved in front of the fiber-head automatically. 
The iodine absorption lines superimposed on the stellar spectra provide 
a stable reference for measuring stellar radial velocities (Marcy \& 
Butler 1992). The cell is a sealed glass cylinder with a diameter of 50 
mm and an optical path of 40 mm. The cell is covered with distributed 
strips of heating tape and mounted in the center of a vacuum vessel. The 
iodine molecules in the cell fully evaporate at 65 $^\circ$C. In order 
to make sure that the iodine molecules fully evaporate without thermal 
broadening of the I$_2$ features, the temperature controller maintains 
the cell at 65 $^\circ$C after a 5 minutes warm-up during our PRV 
observations.

\subsection{Fiber optics}

\begin{figure}
\begin{center}
\includegraphics[angle=0,scale=0.3]{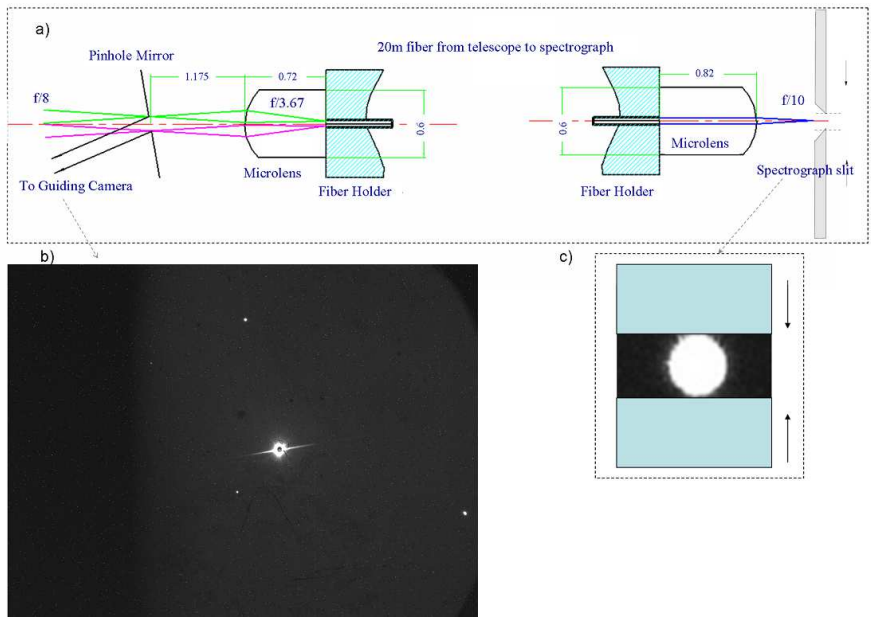}
\caption{a) Fibre feed and exit optics. b) An image acquired by guiding CCD. c) A schematic diagram of the adjustable slit}
\label{FigFiber}
\end{center}
\end{figure}

Using fiber optics to feed the spectrograph, we could mount WES in a 
thermally and mechanically isolated environment to provide an 
exceptionally stable echelle order image on the CCD for exoplanet 
research by the radial velocity method. The focal-ratio degradation 
(FRD) is small from about f/3.0 to about f/7.0, for most popular fibers 
used on astronomy (Ramsey 1988). Behind the pinhole mirror with an 
entrance aperture, a microlens glued on top of the fiber images the 
telescope pupil onto its rear surface with a diameter of 46 $\mu$m 
filling 92\% of the 50 $\mu$m core fibre's entrance-surface, with a 
focal ratio of f/3.67 which is well-suited to minimizing FRD in the 
fibre optics (Fig.~\ref{FigFiber}a). The focal ratio is still 
$\sim$f/3.7 at the other end of the fibre. The microlens at fiber-exit 
produces an f/10 light cone which is the same as the focal-ratio of 
collimator of the spectrograph. An adjustable slit is placed in the 
position of fiber-exit re-imaging. The parameters of the fiber optics 
are summarized in Table~\ref{TabFiber} and shown in 
Fig.~\ref{FigFiber}a.

\begin{table}
\begin{center}
\caption{Fibre optics characteristics}
\label{TabFiber}
\begin{tabular}{lcc}
\hline\hline
Pinhole diameter &  100 $\mu$m    \\ 
Sky aperture     & 2.6 arcsec & \\ 
Fiber core diameter & 50 $\mu$m    \\  \hline
Microlens:		 & Entrance	& Exit \\
\quad Lens diameter &	0.60 mm &	0.60 mm \\
\quad Lens thickness&	0.72 mm &	0.82 mm \\
\quad Radius of curvature	&0.35 mm	&0.35 mm \\
\quad Focal length & 0.38 mm &	0.50 mm\\ 
\quad Lens glass type	 &N-SF66	&KZFS12 \\
\quad Glass refractive index nd	&1.92&	1.65 \\
\quad Output focal ratio  &	f/3.67	&f/10 \\
\hline
\end{tabular}
\end{center}
\end{table}

The fiber coupling has many advantages (Pfeiffer et al 1998). The 
spatial information at the fiber entrance is almost perfectly scrambled 
through the optical fibre. But seeing variations and guiding 
imprecisions mostly result in angular fluctuations which induce some 
variation in the illumination of the echelle grating and thus introduce 
some error in RV measurements (Raskin et al. 2011). And the modal noise 
will reduce the SNR of the spectrum as described by Dr. Grupp (2003). In 
order to remove the fiber noise restrictions and obtain a higher SNR 
spectrum, we installed a mechanical fiber-shaking device in front of 
fiber-exit outside the spectrograph box. The rotating speed of the fiber-shaking device can 
be adjusted by voltage of motor drive for adapting to different exposure 
time.

\subsection{Spectrograph}

\begin{figure}
\begin{center}
\includegraphics[angle=0,scale=0.25]{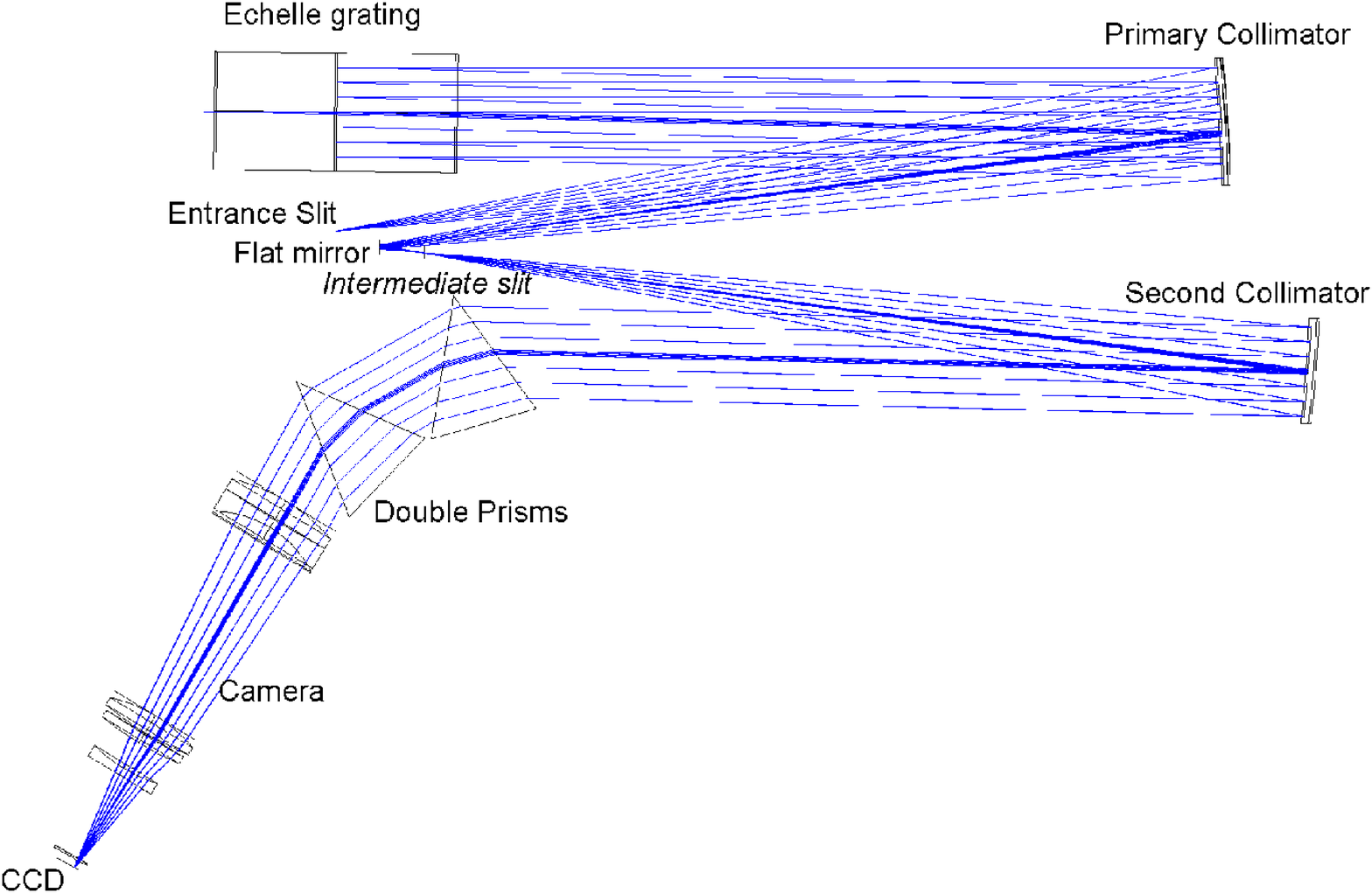}
\caption{The optical layout of WES with collimators, echelle grating, fold mirror, cross-disperser, and camera.}
\label{FigWESlayout}
\end{center}
\end{figure}

The optical layout of WES (Fig.~\ref{FigWESlayout}) is based on a white 
pupil design (Baranne 1988), one of the best solutions for high 
resolution spectrographs, which can be freed easily from scattered light 
produced at the echelle grating, avoids vignetting, and minimizes 
cross-disperser size compared to other designs. The ZEMAX description of the optimised optical system can be
found in Appendix A.

\begin{figure}
\begin{center}
\includegraphics[angle=0,scale=0.2]{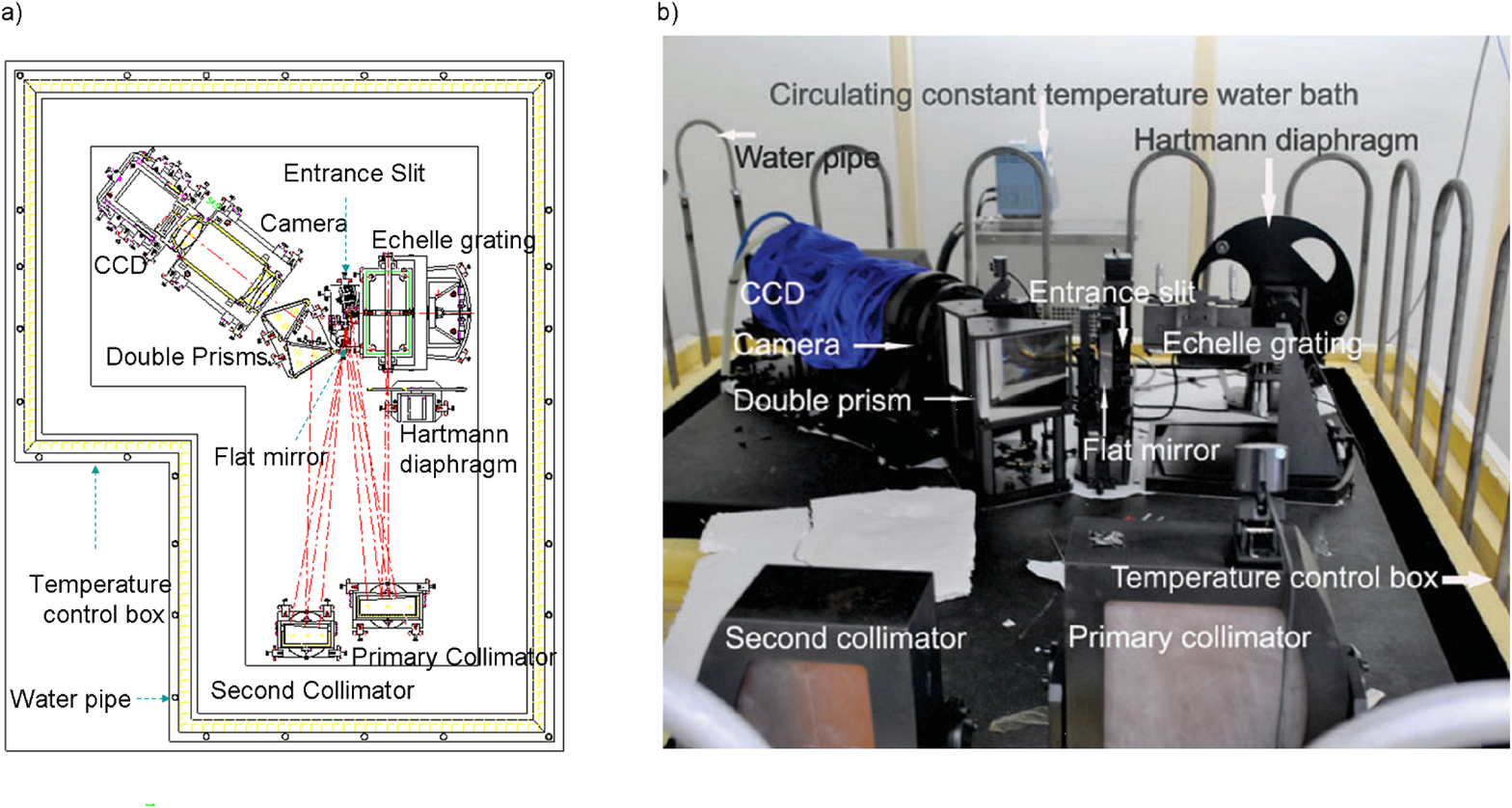}
\caption{a)The layout of WES with the mechanics. b) A photograph of WES mounted on the optical bench for initial testing.}
\label{FigWESphoto}
\end{center}
\end{figure}

\subsubsection{Collimators}

Two collimators of WES were cut from a single parabolic mirror with a 
focal length of 925 mm. The main collimator with a size of 
230$\times$180 mm transforms the f/10 beam to a collimated beam with a 
diameter of 92.5 mm. The beams which are dispersed by echelle grating, 
are again collected by the main collimator and focused at an 
intermediate focal plane. A flat mirror is placed in front of the 
intermediate focal plane in order to fold the light path, which reduces 
the volume of WES and allows an intermediate slit to be placed at the 
intermediate focal plane. Only the light from the intermediate focal 
plane can pass the intermediate slit, stopping most of the stray light 
produced at the echelle grating and roughness of the optical surfaces 
and edges (Kaufer et al. 1999). The second collimator, also called 
transfer collimator, with a size of 230$\times$135 mm, transforms the 
converged beam to a collimated beam cross-dispersed by prisms.  A 
Hartmann diaphragm with left and right semicircle is placed in the 
collimated beam, in front of the echelle grating, for automatic 
focusing.

\subsubsection{Dispersing element}

Because of our compromise between cost and throughput of small 
telescope, the WES uses a R3 echelle grating with blaze angle of 
$\theta=71.5^\circ$ as the main dispersing element. The echelle grating 
was made by Newport company with 31.6 grooves/mm and size of 
128$\times$254 mm. The surface was aluminum-coated with the peak 
efficiency of 73\% at $\sim$630 nm. The echelle grating is mounted 
facing downwards to avoid dust settling on the optical surface. The 
grating is used in quasi-Littrow condition with a slight tilted angle 
($0.7^\circ$).

The white-pupil layout minimizes the size of the cross-disperser. The 
WES uses a pair of two identical LF5 prisms with a base length of 103 
mm, a width of 125 mm, a height of 134 mm, and an apex angle of 
41$^\circ$.

\subsubsection{Camera and CCD}

\begin{figure}
\begin{center}
\includegraphics[angle=0,scale=0.2]{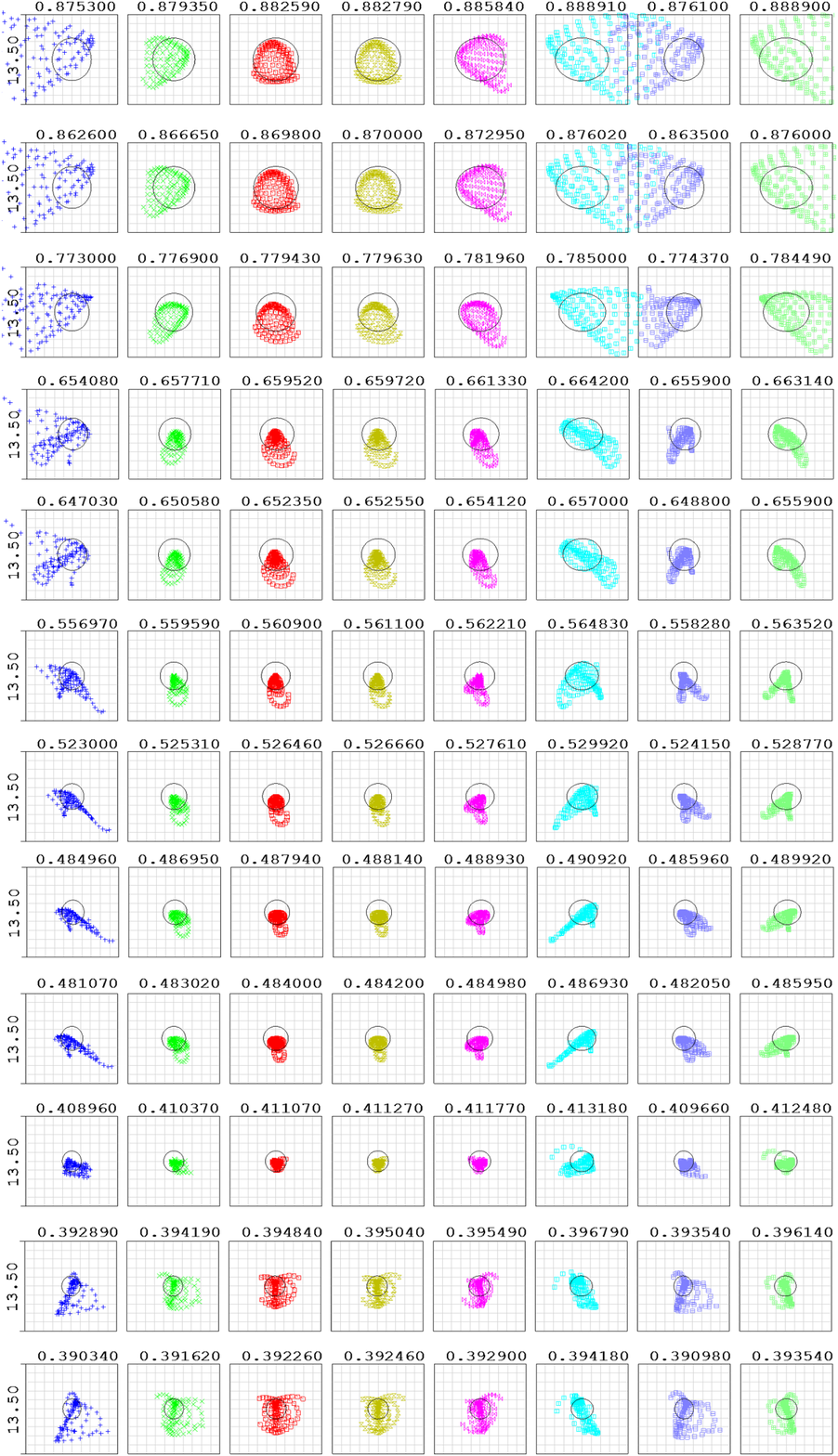}
\caption{ The spot diagrams across the echellogram for the complete optical system of WES design. Wavelength is indicated in $\mu$m above each 1 $\times$ 1 CCD pixels box (13.5 $\mu$m). From bottom left to up right is blue to red. The solid line implies Airy disk.}
\label{FigSpot}
\end{center}
\end{figure}

The f/3.0 dioptric camera with a diameter of 116mm images the echellogram 
onto a 2K$\times$2K Andor iKon-L DZ936N-BV CCD with a 13.5 $\mu$m/pixel 
scale. The image quality of the camera designed for WES is that 
calculated 80\% encircled energy spot diameter is smaller than 25 
$\mu$m. The spot diagrams across the echellogram for the complete 
optical system of WES design is shown in Fig.~\ref{FigSpot}. The free 
spectral range of WES with 100 spectral orders from 370 nm to 976 nm 
simulated on the a 2048 $\times$ 2048 13.5 $\mu$m pixels CCD are shown 
in Fig.~\ref{FigRewdata}a. The arrangement of echelle grating and cross 
disperser results in a line tilt relative to the echelle dispersion 
direction. Although the geometric effect of line tilt is to degrade the 
resolving power, this effect is negligible at all but the largest tilts 
(Hearnshaw et al. 2002). The R3 echelle grating of WES is used in 
quasi-Littrow condition with a slight tilted angle ($0.7^\circ$) causing 
4.2$^\circ$ line tilt. Then the whole spectrum is not slightly tilted to 
align the slit images with the CCD rows specially because the spectral 
lines tilt little. The impact of slit tilt and anamorphism on 
spectrograph resolution is negligible as shown in Fig.~\ref{FigTHAR}a 
and Fig.~\ref{FigRewdata}b. In order to keep the camera in a very stable 
mechanically condition for measuring PRV, the symmetrical mountings of 
camera are manufactured in the shape of circular tube because of its 
excellent resistance to gravity deformation. The whole barrel of camera 
is installed on an electric guide rail for automatic focusing. The CCD 
operates at 0.05MHz with a readout noise of 3.6 e$^{-}$. The temperature 
of the chip is set to -95 $^\circ$C by water cooling which is good for 
low dark current and environment stability.

\subsection{Environment Control}

The optical elements of WES are all installed on a marble 
vibration-proof optical bench which is 1500 kilograms with a thickness 
of 20 cm (see Fig.~\ref{FigWESphoto}b). There is an air compressor for 
an air-supported vibration isolation system under the optical bench for 
avoiding mechanical vibration for frequencies below 3 Hz. Temperature 
and atmospheric pressure controlled for the long-term are the most 
important parameters for environment stability. This bench is placed in 
an air-conditioned room to limit temperature and humidity changes. 
Unlike the HARPS spectrograph, which is operated in vacuum (Pepe et al. 
2000) and the HERMES which was atmospheric pressure controlled (Raskin 
et al. 2011), WES is just operated in a temperature controlled 
environment at present (see Fig.~\ref{FigWESlayout}). The design of 
atmospheric pressure control for WES will be developed in the near 
future.

The optical bench is wrapped by a laminated box made of two panels of 
polyurethane. There is a circulatory pipe with constant temperature 
water and heat reflecting plate made of aluminum between two panels. The 
CCD was also wrapped by a polyurethane box to isolate thermal 
fluctuations. There are two refrigerated and heating circulators outside 
the air-conditioned room, one for the spectrograph box setting 
temperature to 20 $^\circ$C and the other for CCD cooling setting 
temperature to 18 $^\circ$C. When using water cooling, the internal fan 
of the CCD is switched off via the software. The sensors for 
temperature, humidity, and pressure measurements are placed inside and 
outside the temperature control box. All of the mounts for the large 
optical elements (camera, collimators, echelle grating, cross 
prisms, etc) are made of Invar with very small coefficient of thermal 
expansion (CTE), 1.6$\times$${10}^{-6}$/$^\circ$C. The mountings of 
camera lens are manufactured from different cast iron with different 
CTE, which matches the CTE of the lens. The symmetry of the structure 
minimizes gravity deformation. All the elements are locked and shut off 
power after adjustment.

\section{Observations and data reduction}

We have two observation modes: the normal mode and the PRV mode. In the 
normal mode, we observe the object at the required spectral resolutions 
(40,000-55,000) without iodine cell, providing data for common stellar 
physics research.  In the PRV mode, we obtain the spectrum at the 
spectral resolution of R$\sim$50,000 with iodine cell.  We observe a 
template iodine-free stellar spectrum at higher resolution 
(R$\sim$55,000).  We always arrange the template observations with high 
quality at very clear nights. Then we also observe a rapidly rotating B 
star with the iodine cell in order to calculate the instrumental profile 
at the same night. The template observations can also provide data for 
studies of stellar abundance, gravity gradient, and other physical 
parameters, which is crucial for the formation and evolution of 
planetary systems. We use the same method of RV analysis as described by 
Sato et al. (2002).

We chose some RV standard stars ($\epsilon$ Vir, $\iota$ Per, $\tau$ 
Cet, etc) for regular observations to test the stability of the 
spectrograph and detect the instrumental system drifts. We also observe 
some targets of known exoplanet hosts ($\beta$ Gem, $\upsilon$ And, etc) 
for further monitoring to derive accurate parameters of planetary 
systems and reveal additional planets. We have obtained 106 frames in 25 
nights distributed during more than 4 years for $\beta$ Gem (Hatzes et 
al. 2006). More than 90 objects for RV measurement with V magnitude from 
0 to 6 have been observed since 2010 after WES installed.

The raw data reduction is made using standard procedures in the 
{\tt{echelle}} package in Image Reduction and Analysis Facility 
(IRAF\footnote{IRAF is distributed by the National Optical Astronomy 
Observatories, which are operated by the Association of Universities for 
Research in Astronomy (AURA) under cooperative agreement with the 
National Science Foundation.}).  The reduction procedure is summarized 
as follows:
\begin{itemize}
\setlength{\itemsep}{0pt}
\setlength{\parsep}{0pt}
\setlength{\parskip}{0pt}
\item[-]  Combination of bias and bias subtraction for the frames of objects, flat fields, ThAr.
\item[-]  Combining and mosaicing flat fields. Because of the different throughput in the blue and red part of the spectrograph, we acquire flat fields with three different exposure times and combine them respectively. We trim different regions of three different types of flat fields, and then mosaic into one flat field.
\item[-] Normalizing the flat field and flat fielding the data.
\item[-] Background subtraction. Unlike the long-slit spectrograph, both the sky and the target's light are recorded at the same location on the CCD with the fiber-fed spectrograph. We should observe the sky spectra separately. The level of scattered light is very low with the layout of white pupil. We can do or skip this step according to our scientific purposes.
\item[-] Extraction of the spectra and wavelength calibration.
\item[-] Continuum normalization. In the observations with iodine cell, it is difficult to fit the continuum because of iodine absorption spectra superimposed on the stellar spectra. We normalize the stellar spectra divided by the spectra of combined flat field. 
\end{itemize}

\section{Performances}

\subsection{Wavelength range and Spectral resolution}

For a fiber-fed spectrograph without FRD effect, the nominal spectral 
resolution $R=\lambda/\Delta\lambda$ of WES with a collimated beam size 
of 92.5 mm, a sky aperture of 2.6 arcsec, and a R3 echelle grating at 
the 1.0 m WHOT is about 44 000. Each image of fiber core is sampled by 
3.0 $\times$ 3.0 pixels on the CCD. The resolution can be increased to 
57000 with a minimum of 2.2 pixels sampling by narrowing the adjustable 
slit at the fiber-exit.

\begin{figure}
\begin{center}
\includegraphics[angle=0,scale=0.25]{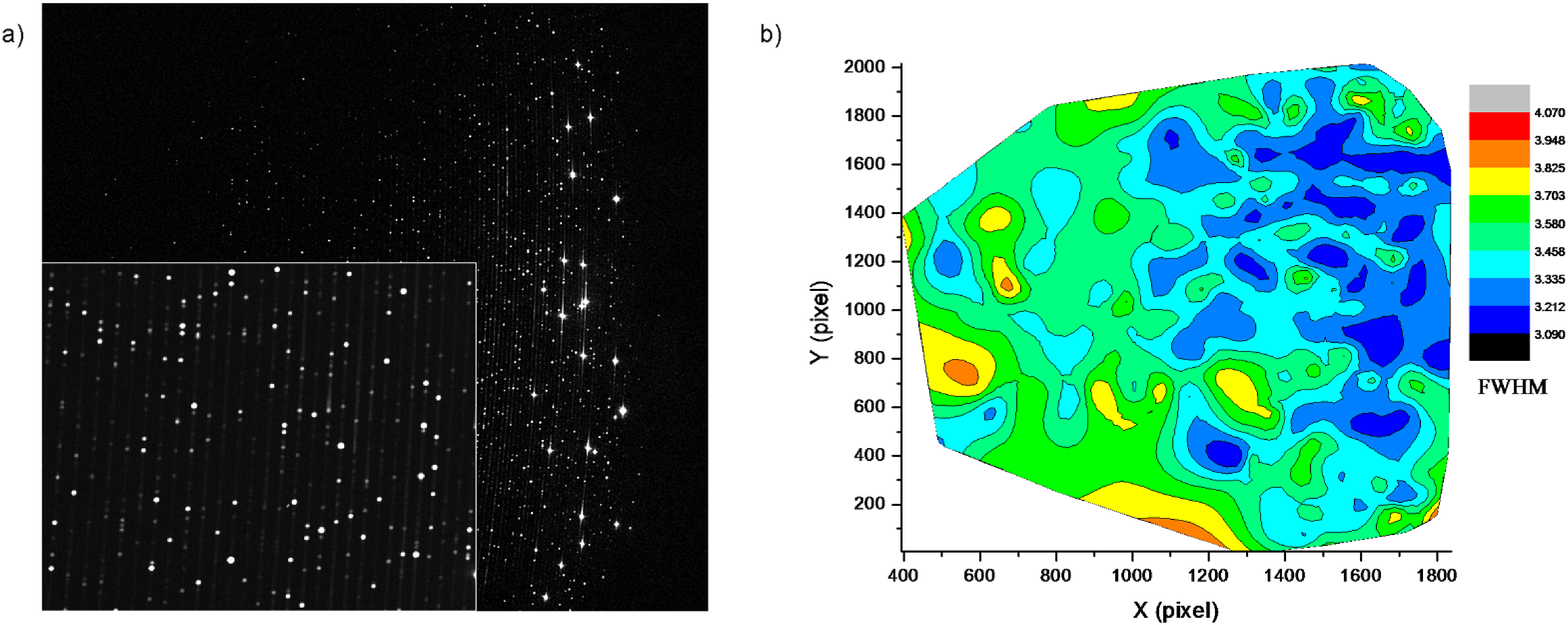}
\caption{a) The full-frame raw data of ThAr. At bottom left the inset shows a magnified section of the echelle image. b) The contour of FWHM measured from the ThAr frame.}
\label{FigTHAR}
\end{center}
\end{figure}

The image of the ThAr frame with full slit width is shown in 
Fig.~\ref{FigTHAR}a. We measured the full width at half maximum (FWHM) 
of the image of fiber core all through the CCD to examine the overall 
image quality of WES. The FWHMs of all single emission lines are chosen 
and their contour is shown in Fig.~\ref{FigTHAR}b. The image quality at 
the focal plane is good and uniform, in the sense that the actual 
average FWHM of the image of fiber core is 3.38 pixels, and the 
variation is less than 0.86 pixels. We note that the left region has a 
slightly larger FWHM than the right, which is possibly caused by the 
measurement accuracy effected by low SNR or a misalignment of the CCD 
camera with respect to the optical axis. We will consider it in the 
future adjustment.

\begin{figure}
\begin{center}
\includegraphics[angle=0,scale=0.25]{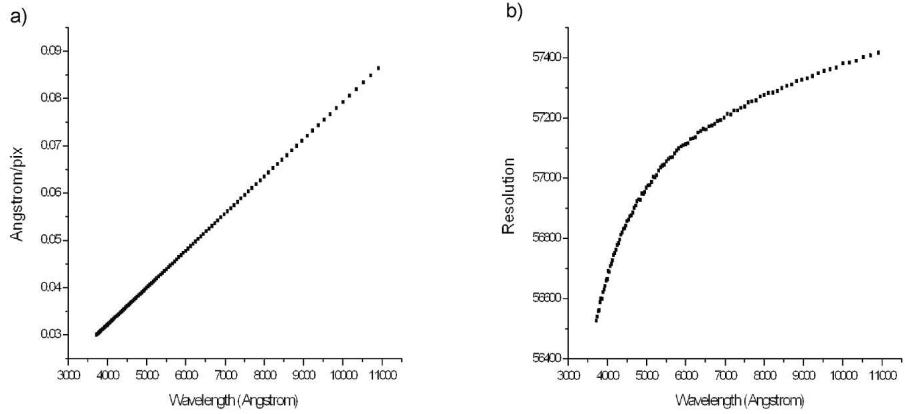}
\caption{a) The spectral sampling per pixel of each order according to different wavelengths. b) The spectral resolution with 2.2 pixels sampling.}
\label{FigSampPix}
\end{center}
\end{figure}

The arc spectra allows us to get a dispersion solution (wavelength 
versus pixel and order number) of WES. We identify the emission lines 
using the task of ``ecidentify'' in the echelle package of IRAF with 
four-order Chebyshev function after tracing and extracting a ThAr frame. 
The RMS of the identification is 0.003 \AA \ (according to $\sim$0.06 
pixels, $\sim$147 m/s).  The wavelength range covered by each order and 
the corresponding sampling ratio per pixel can be derived after the 
identification (see Fig.~\ref{FigSampPix}a). The spectral resolution is 
defined as $R=\lambda/\Delta\lambda$. We derived the $\Delta\lambda$ 
from the FWHM of the ThAr emission lines projected alone spectral 
direction after flux integral along cross-dispersion direction.

\begin{figure}
\begin{center}
\includegraphics[angle=0,scale=0.25]{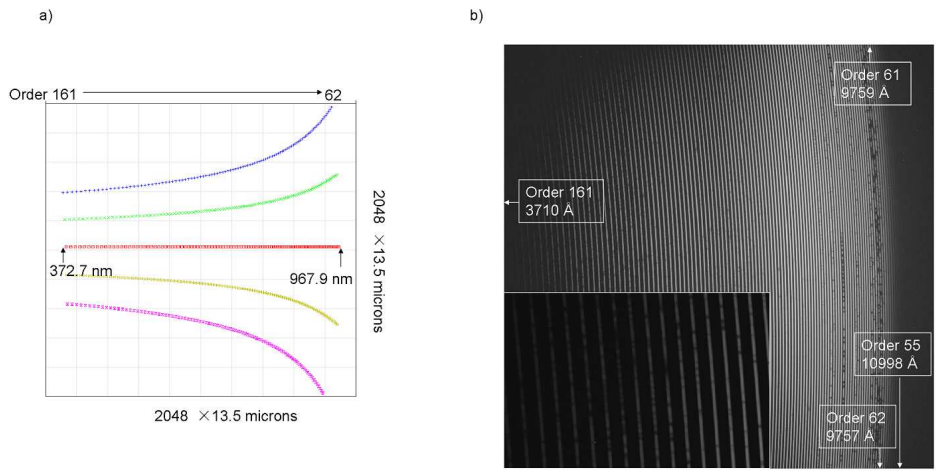}
\caption{a) The free spectral range of WES with 100 orders simulated on the 27.648 $\times$ 27.648 mm CCD, blue is towards the bottom
left and red is towards the top right. b) The full-frame raw data of $\tau$ Cet displaying 107 echelle orders of curved echellogram ranging from 3710 to 10998 \AA. At bottom left the inset shows a magnified section of the echelle image.}
\label{FigRewdata}
\end{center}
\end{figure}

The full-frame raw data obtained from $\tau$ Cet shows the position of 
the curved echellogram on the CCD in Fig.~\ref{FigRewdata}b. The 
rectangular 2K$\times$2K CCD with 13.5$\mu$m per pixel can image 100 
orders of continuous spectrum from 371 to 976 nm and 7 further extended 
orders of slightly truncated spectrum from 976 to 1,100 nm. The spectral 
resolution power R=$\lambda/\Delta\lambda$ is 40,600 with full 
slit-width at 550 nm. The maximum spectral resolution is 57,000 with a 
2.2 pixels sampling at 550 nm (see Fig.~\ref{FigSampPix}b). We set the 
throughput of full slit-width as 1.0. The spectral resolution and 
relative throughput of different slit-widths measured through ThAr 
frames are shown in Table~\ref{TabSlit}. Because we can not measure the 
slit-width directly, the values of slit-width in Table~\ref{TabSlit} is converted by 
the software according to the step of the drive motor and it is the 
relative value (Coefficient $\times$ Physical width $+$ Constant), not the physical 
value. So we mainly focus on the resolution and the relative throughput.

\begin{table}
\begin{center}
\caption{The measured spectral resolution and throughput of different slit-widths.}
\label{TabSlit}
\begin{tabular}{lccccc}
\hline\hline
Slit-width$^a$ & Relative throughput & FWHM  &  Spectral Resolution  \\ 
($\mu$m) &     & (pixels) & (R) \\ \hline
135	& 1.00 &	3.09	& 40600 \\ 
130	& 0.98  &	3.01	& 41700 \\
125	& 0.96 &	3.00 & 41800 \\
120	& 0.93 &	2.93	& 42900 \\
115  & 0.91 & 	2.90	& 43300 \\
110	& 0.87 &	2.83	& 44500 \\ 
105	& 0.83 &	2.75	& 45700 \\
100	& 0.77 &	2.69 & 46700 \\
95	& 0.74 &	2.55	& 49200 \\
90  & 0.67 & 	2.49	& 50400 \\
85  & 0.60 & 	2.38	& 52900 \\
80	& 0.53 &	2.29	& 54900 \\ 
\hline
\end{tabular}
\end{center}
{\it Note: $^a$The values is the relative value (Coefficient $\times$ Physical width $+$ Constant), not the physical value.}
\end{table}

\begin{figure}
\begin{center}
\includegraphics[angle=0,scale=0.25]{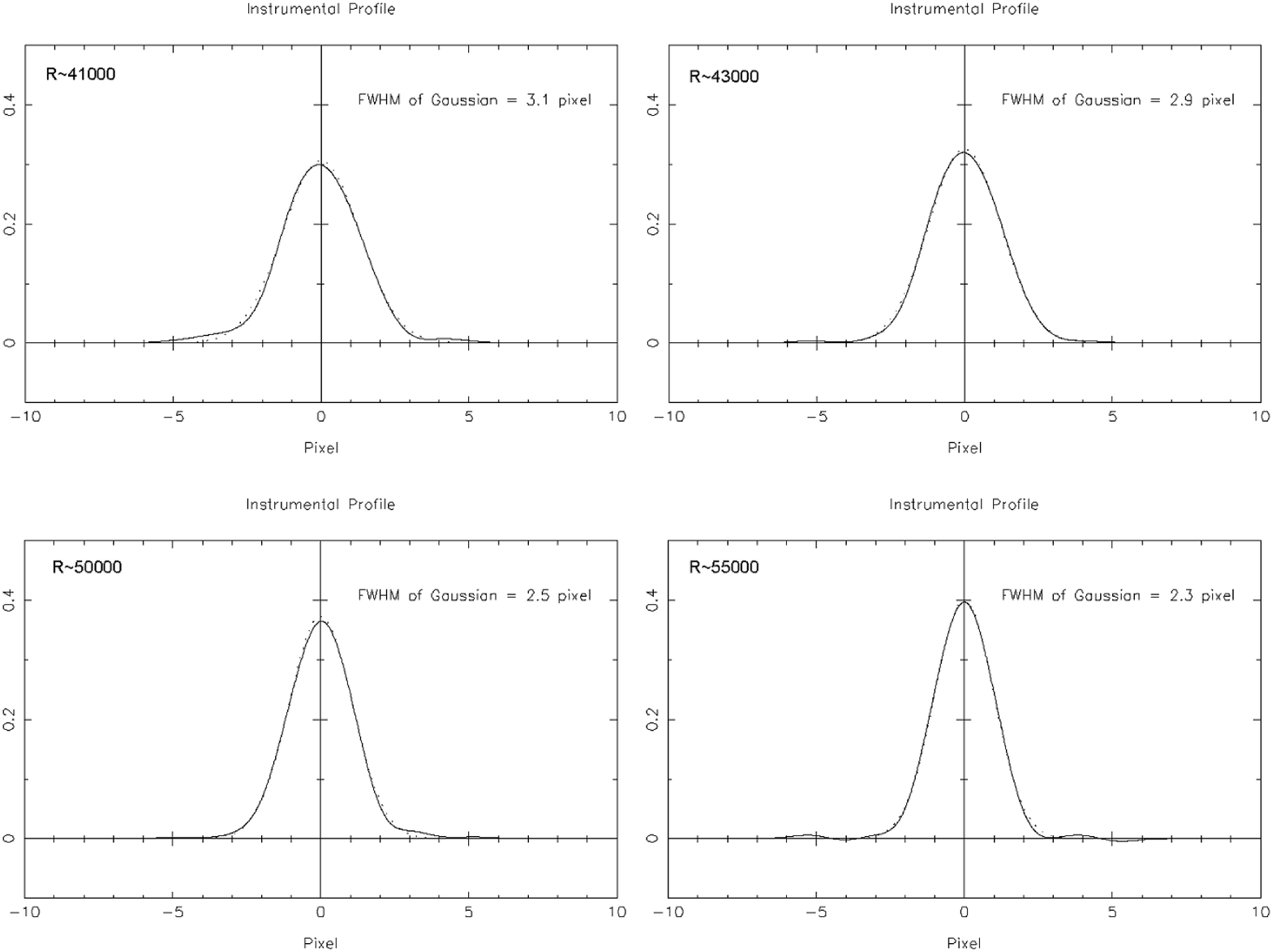}
\caption{The IPs of WES with different resolutions, at the center of the 108$^{th}$ blaze order ( $\sim$5560 $\AA$) along the wavelength (CCD line), derived from B-star + I$_{2}$ spectra. The solid and dashed line are the IP from multi and single Gaussian fitting respectively in this data modeling technique.}
\label{FigIP}
\end{center}
\end{figure}

The IPs of WES with different resolutions, at the center of the 
108$^{th}$ blaze order ($\sim$5560 $\AA$) along the wavelength (CCD 
line), was modelled through the B-star + I$_{2}$ spectrum (Butler et al. 
1996; Endl et al. 2000, Sato et al. 2002) as shown in Fig.~\ref{FigIP}.  
We modelled the IP with a combination of a central and ten satellite 
Gaussian profiles which are placed at appropriate intervals and have 
suitable widths. The IP of WES is symmetric and close to the Gaussian 
profile nearly, which is very important for the PRV measurement (Kambe 
et al. 2002).

The cross-disperser provides a minimum order separation of 11 pixels in 
the reddest part of the spectrum, and a maximum order separation of 26 
pixels in the bluest part of the spectrum. Because of more than 3 pixels 
of FWHM in each order and the possible spectra overlapping redward of 
6,000 \AA, it is not good enough to use simultaneous thorium technique 
(Baranne et al. 1996).

\begin{figure}
\begin{center}
\includegraphics[angle=0,scale=0.3]{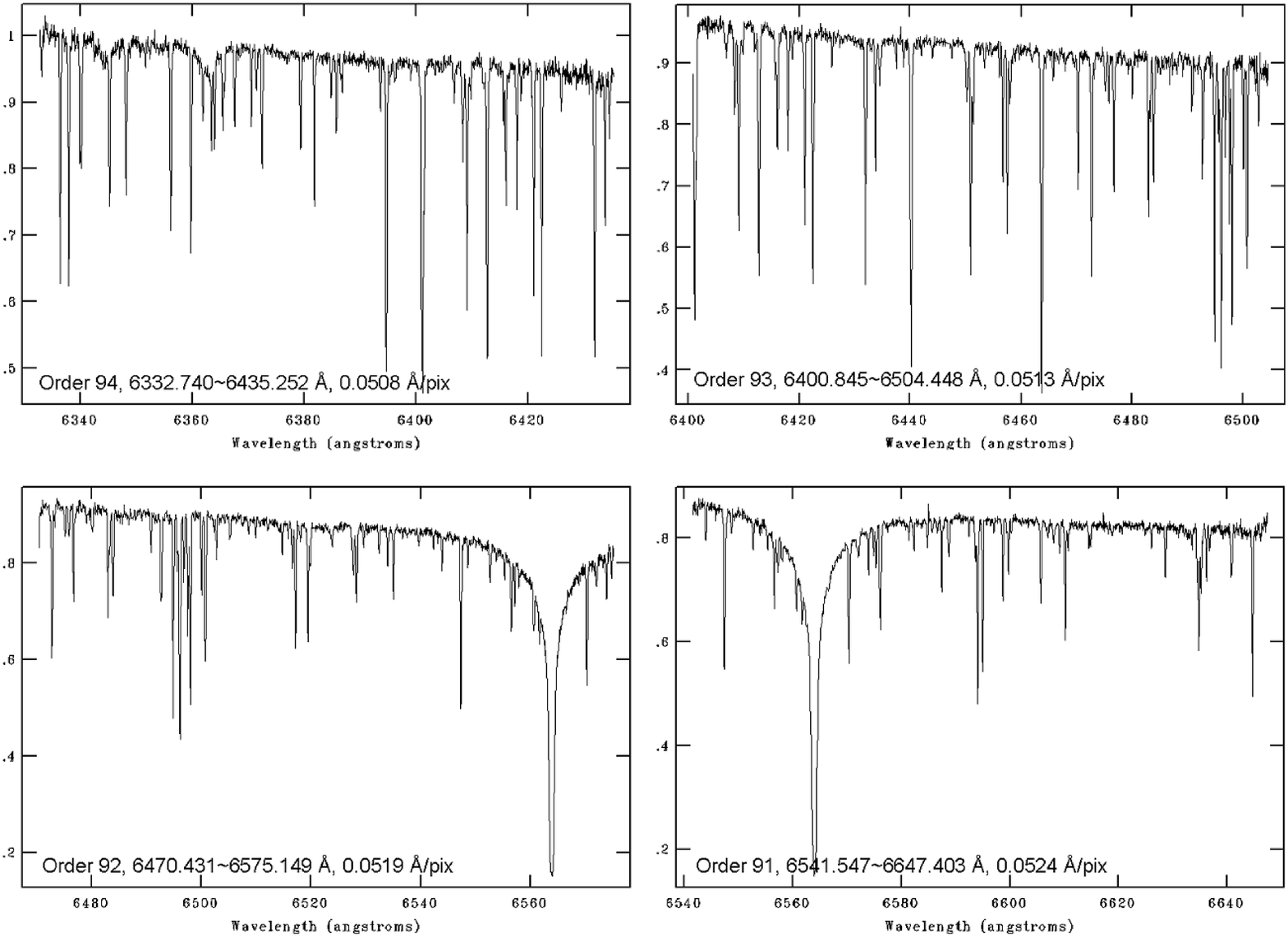}
\caption{Some blaze orders of the normalized spectra of  $\iota$ Per (Vmag=4.05), obtained with 30 min exposure time at R = 50,000.}
\label{FigSpectra}
\end{center}
\end{figure}

Fig.~\ref{FigSpectra} shows the spectra of some blaze orders of $\iota$ 
Per (HD19733, Spectral Type: F9.5\uppercase\expandafter{\romannumeral5}, 
Vmag=4.05) observed with 30 min exposure time at R=50,000. The spectra 
is reduced after continuum normalization.


\subsection{Efficiency}

The system efficiency is the ratio of photons incident at the top of the 
atmosphere at zenith and the detected photons on the spectrograph CCD 
(Aceituno et al. 2013). This efficiency includes atmospheric 
transparency, throughput of all optical elements, and guiding losses.

We estimated the efficiency of the whole spectroscopic system using as 
much as possible the specifications given by the corresponding 
instrument data sheets and the reflectivity or transmissivity of each 
element we measured.  Before the efficiency test, we measured the 
reflectivity or transmissivity of each element using a spectrophotometer 
CM-2300d (Konica Minolta, Japan), after we cleaned the optical elements. 
The efficiency of each element is listed in Table~\ref{TabElement}. The 
mean atmospheric transmission is 0.74 in the V band tested by 
photometric system of WHOT (Hu et al. 2014). The efficiency of 1-m 
telescope is 0.67, degraded by the main telescope mirror reflections 
roughly 0.91, the secondary telescope mirror reflections 0.84 and 0.87 
pass secondary mirror block.  The reflectance of the pick-off mirror is 
0.80. The throughput of 0.1 mm guiding pinhole is 0.80 at the median seeing 
of 1.7$''$, as described in Sect. 2.1. The throughput of 50 $\mu$m fiber 
optics is 0.58 tested by NIAOT. The efficiency of the spectrograph 
itself was originally estimated to be around 0.30 in the V band, using 
the specifications given by data sheets (Table~\ref{TabElement}). So the 
overall expected efficiency is 0.74*0.67*0.8*0.8*0.58*0.30=0.055 in the 
V band, including atmospheric transparency and guiding pinhole losses. 
The efficiency will be higher at 620nm in the R band.
 
\begin{table}
\begin{center}
\caption{The efficiency of each element}
\label{TabElement}
\begin{tabular}{llcl}
\hline\hline
Element & Type  & Efficiency  &  origin  \\ \hline
Atmospheric transmission & Transmission& 0.74 &	Measured  \\ 
Telescope & & & \\
\quad Main mirror &	Mirror reflectivity&	0.91&	Measured \\
\quad Secondary mirror&	Mirror reflectivity&	0.84&	Measured \\
\quad &Secondary mirror block pass&	0.87	&Data sheet \\ 
Pick-off mirror	&Mirror reflectivity&	0.80	&Measured \\ 
Pinhole mirror&  	Throughput at the seeing of 1.7$''$& 	0.80&	Calculated \\ 
Fiber optics&	Throughput of fiber optics&	0.58&	Measured \\
Spectrograph & & & \\
\quad Two collimators &	Mirror reflectivity &	0.945 &	Data sheet\\
\quad Fold mirror &	Mirror reflectivity &	0.986	 &Data sheet\\
\quad Prism	 & & & \\
 &Total transmission &	0.88 &	Data sheet\\
 &Total entrance/exit transmission &	0.916	 &Data sheet\\
\quad Echelle grating&  & & \\	
 &Overfilling	&0.89	&Data sheet\\
	&Absolute efficiency at 630nm&	0.73& 	Data sheet\\
\quad Camera	&All lenses total transmission&	0.81&	Data sheet\\
\quad CCD & & & \\	
&Quantum efficiency&	0.89	&Data sheet\\
	&Entrance/Exit transmission&	0.93	&Data sheet\\

\hline
\end{tabular}
\end{center}
\end{table}

The throughput of the photometric system in each band at WHOT has been 
estimated by observing a series of standard stars on photometric nights. 
We arranged efficiency comparison tests between WES and photometric 
system through dome-flat at the same night. The brightness of the flat 
field lamp is stable which eliminate the effect of weather and seeing. 
The specific steps of dome-flat test are as follows:

\begin{itemize}
\setlength{\itemsep}{0pt}
\setlength{\parsep}{0pt}
\setlength{\parskip}{0pt}
\item[1)] In the night without the Moon, the dome was closed completely after dark.
\item[2)] Open the flat-field lamp half an hour, waiting for the light becomes stable. Make sure that the telescope is not moving.
\item[3)] Acquire the dome-flat field through the photometric system. In order to avoid shutter effect, the exposure time should not be too short.
\item[4)] Acquire the dome-flat field through the WES
\item[5)] Data analysis, specific as follows:
	\begin{itemize}
	\setlength{\itemsep}{0pt}
	\setlength{\parsep}{0pt}
	\setlength{\parskip}{0pt}
	\item[i)] Calculate the photon number per square centimeter per second (cm$^{-2}$s$^{-1}$) obtained from a filter of the photometric system ($N_{phot}$). And extract the spectra of dome-flat obtained from WES.
	\item[ii)] The dome-flat spectrum of WES is convolved with the transmittance curve of the filter. Then the photon number in this band is obtained from WES, ($N_{spec}$), translated to per square centimeter per second according to the area of the pinhole aperture.
	\item[iii)] Through the efficiency of the photometric system, ($E_{phot}$), we can get the efficiency of WES, $E_{spec} = E_{phot}\times( N_{spec}/ N_{phot} )$.
	\end{itemize}	
\end{itemize}

The dome-flat test shows that the efficiency of WES is 38.5\% of 
photometric system with efficiency 0.183 in V band including atmospheric 
transparency in October 2010. So the efficiency of WES is 
0.385*0.183$\approx$0.070 in the V band without guiding losses. 
Considering throughput of 0.1 mm guiding pinhole for the point source, 
the final efficiency is 0.07*0.80=0.056 including atmospheric 
transparency and guiding loss. It is similar to the value estimated 
using data sheets and the reflectivity of each element as described 
above.
 
Table~\ref{TabObs} shows the SNR and efficiency computed using some 
existing scientific observation data. We acquired a spectrum of HD19445 
(Vmag=8.06) in an hour exposure with an SNR of 106 at 550 nm, with the 
resolution of R$\approx$40,000 and a seeing of 1.4$''$ on November 17, 
2010.
 
\begin{figure}
\begin{center}
\includegraphics[angle=0,scale=0.3]{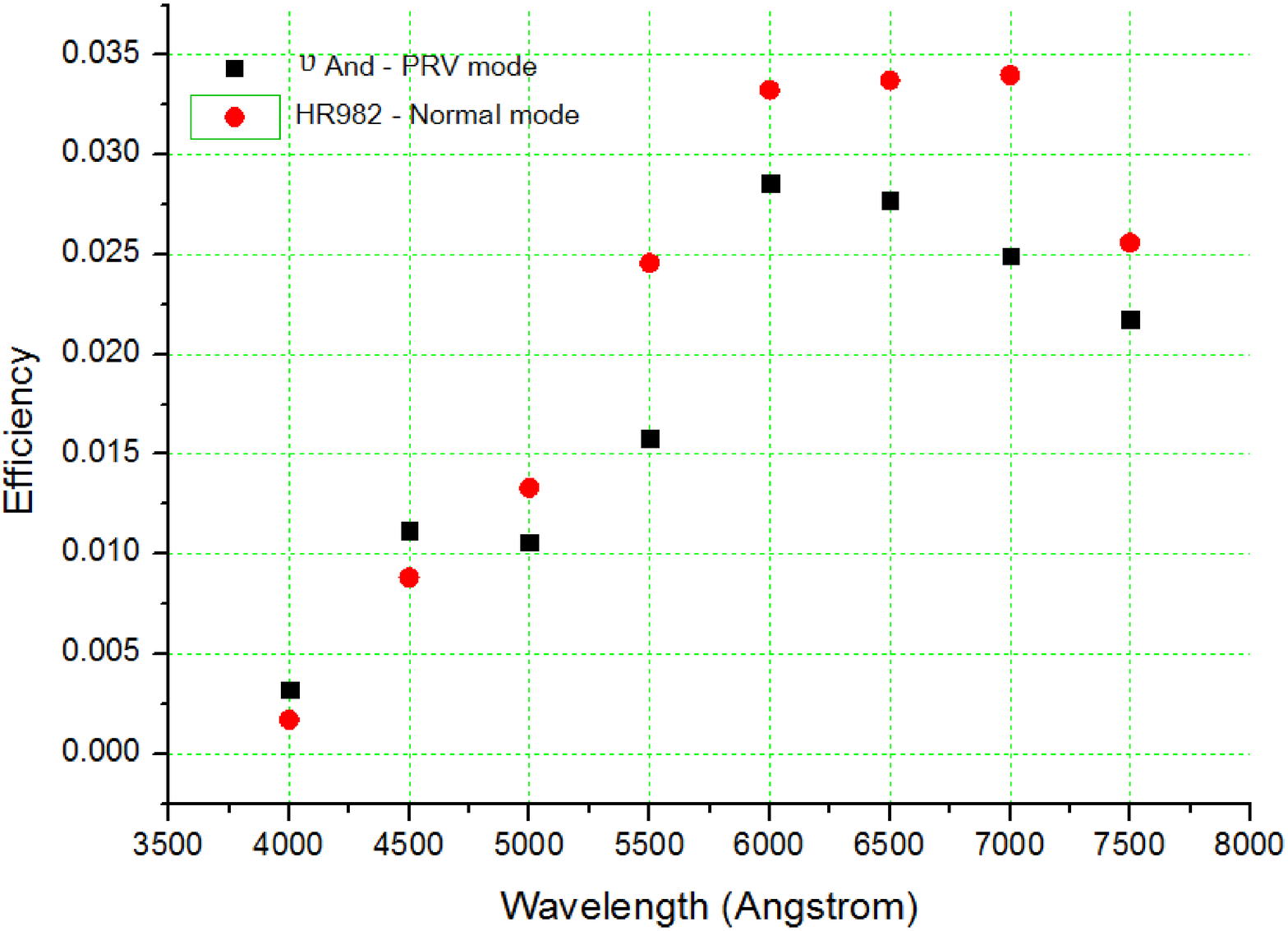}
\caption{ Overall WES efficiency derived from the absolute flux of standard star HR982 and  $\upsilon$ And, respectively corresponding to the normal mode and the PRV mode. }
\label{FigEff}
\end{center}
\end{figure}

In the PRV mode, the throughput of I$_{2}$ cell is 0.77 at 550nm, 0.86 
at 620 nm. The throughput of slit width is $\sim$0.67 at R=50,000. So 
the efficiency in the PRV mode is 0.77*0.67$\approx$0.52 of that in 
normal mode without the I$_{2}$ cell and slit at 550nm. The SNR of 
$\upsilon$ And (Vmag=4.09) is 269 with an exposure of 600 second. 
Through this observation of $\upsilon$ And, the SNR is estimated to be 
112 for Vmag=6 star in 10-min exposure time. The absolute spectral flux 
(photons cm$^{-2}$s$^{-1}$\AA$^{-1}$) of standard stars can be found on 
the website (http://deep-red.sr.unh.edu/starflux/).  We calculated the 
efficiency of two observation modes using observations of HR982 and 
$\upsilon$ And at different wavelengths as shown in Fig.~\ref{FigEff}.

\begin{table}
\begin{center}
\caption{Some observations of WES.}
\label{TabObs}
\scriptsize
\begin{tabular}{lccc|ccc|ccc}
\hline\hline
Object & Exptime & I$_{2}$ cell & R & Vmag & SNR$_{550nm}$ & Efficiency & Rmag & SNR$_{620nm}$ & Efficiency  \\
       &(s)	&(Yes/No)	&	&    &         &at 550nm	&     &	&at 620nm 	\\ \hline
Vega		&15	&No	&50000	&0.03&	280	&0.016	&0.07	&313	&0.032 \\
Deneb		&60	&No	&40000	&1.25&	383	&0.022	&1.14	&415	&0.037 \\
HIP677	&60	&No	&40000	&2.06&	304	&0.029	&2.09	&322	&0.054 \\
HIP96516	&1200	&No	&40000	&5.70&      228	&0.024	&5.56	&307	&0.059 \\
HD19445	&3600	&No	&40000	&8.06&	106	&0.017	&7.60	&130	&0.032 \\
HD115515	&1800	&No	&40000	&9.45&	48	&0.021	&9.79	&55	&0.060 \\
\hline
HD71369	&360	&Yes	&50000	&3.42&	227	&0.010	&2.76	&310	&0.016 \\
 $\upsilon$ And	&600	&Yes	&50000	&4.09&	269	&0.015	&3.64	&360	&0.028 \\
 $\tau$ Cet	&600	&Yes	&50000	&3.50&	307	&0.011	&2.88	&472	&0.023 \\
\hline
\end{tabular}
\end{center}
\end{table}

The best efficiency that already appeared is 2.9\% at 550nm, which is 
lower than expected. Because of no cleaning of the optical instruments 
for a long time, suboptimal seeing, imperfect weather and telescope 
guiding for observation, the efficiency cannot reach expected value.

\subsection{Stray light}

The stray light is primarily reduced by the intermediate slit in front 
of the flat mirror, as described in Sect 2.3.1. The distribution of 
stray light is very local, and became relevant when the signal is weak 
at the blue part of focal plane. We determine the stray light level 
scattered into a pixel, by the ratio of spectral intensity in the 
inter-order to the peak flux in the adjacent orders with the frames of 
flat-field. The highest stray light level is 5.3\% at blue part of the 
focal plane, and less than 0.7\% at most of red focal plane.
	
\subsection{Stability and RV precision}

During the commissioning periods without the most stable conditions, the temperature in the spectrograph box 
changed 1.7 $^\circ$C in six months from 
September 2010 to February 2011, causing 7.3-pixel shift in 
cross-dispersion direction and 0.8-pixel shift in dispersion direction 
on the CCD. The spectrograph is thermally stabilised to $\pm$0.03 
$^\circ$C in one day, $\pm$0.04 $^\circ$C in one week after the 
commissioning.

The RV method for exoplanet research requires very high precision. For 
example, Jupiter imparts a velocity of 12.47 m/s on the Sun. If we want 
to determine a Doppler shift of 10 m/s at the optical band, the 
wavelength shift at 5,000 {\AA} is about 0.00016 $\AA$ which correspond 
to 0.004 pixels on the CCD of WES. So RV observations have high demands 
on spectrograph stability.

\begin{figure}
\begin{center}
\includegraphics[angle=0,scale=0.28]{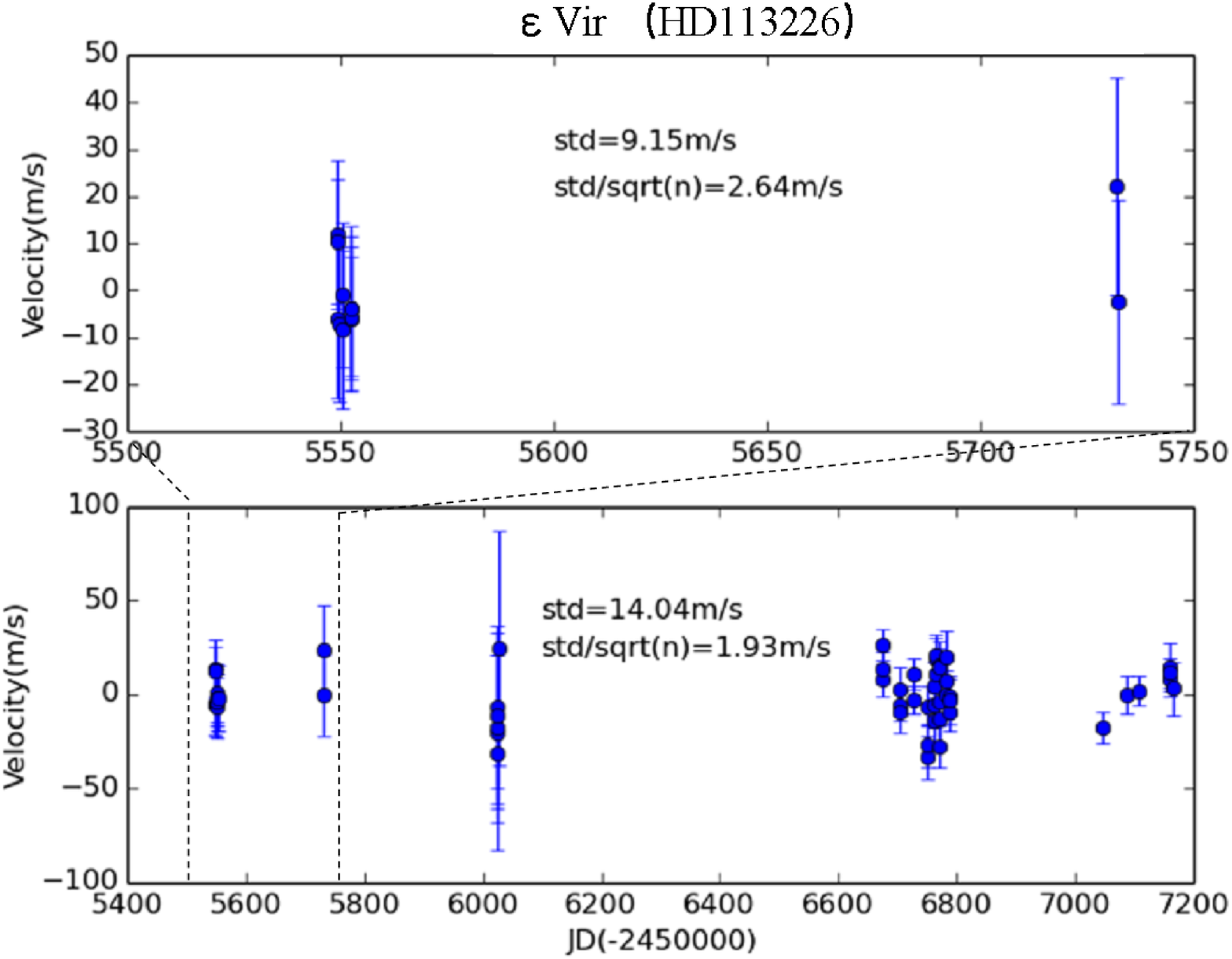}
\caption{The radial velocities of $\epsilon$ Vir (HD113226) obtained from WES. }
\label{FigHD113226}
\end{center}
\end{figure}

The radial velocities of $\epsilon$ Vir (HD\,113226, Vmag=2.79, 
G8\uppercase\expandafter{\romannumeral3}) calculated from the 
observations of WES are shown in Fig.~\ref{FigHD113226} in order to test 
measurement precision of radial velocity. The standard deviation (as 
``std'' shown in the figure) of the measured radial velocities is 9.15 
m/s over a 6-month period (183 days), with the internal measurement 
error (as ``std/sqrt(n)'' shown in the figure) of 2.64 m/s. As shown in 
bottom panel of Fig.~\ref{FigHD113226}, the standard deviation is 14.04 
m/s over a 54-month period (1,617 day, 4.4 years), with the internal 
measurement error of 1.93 m/s.

\begin{figure}
\begin{center}
\includegraphics[angle=0,scale=0.28]{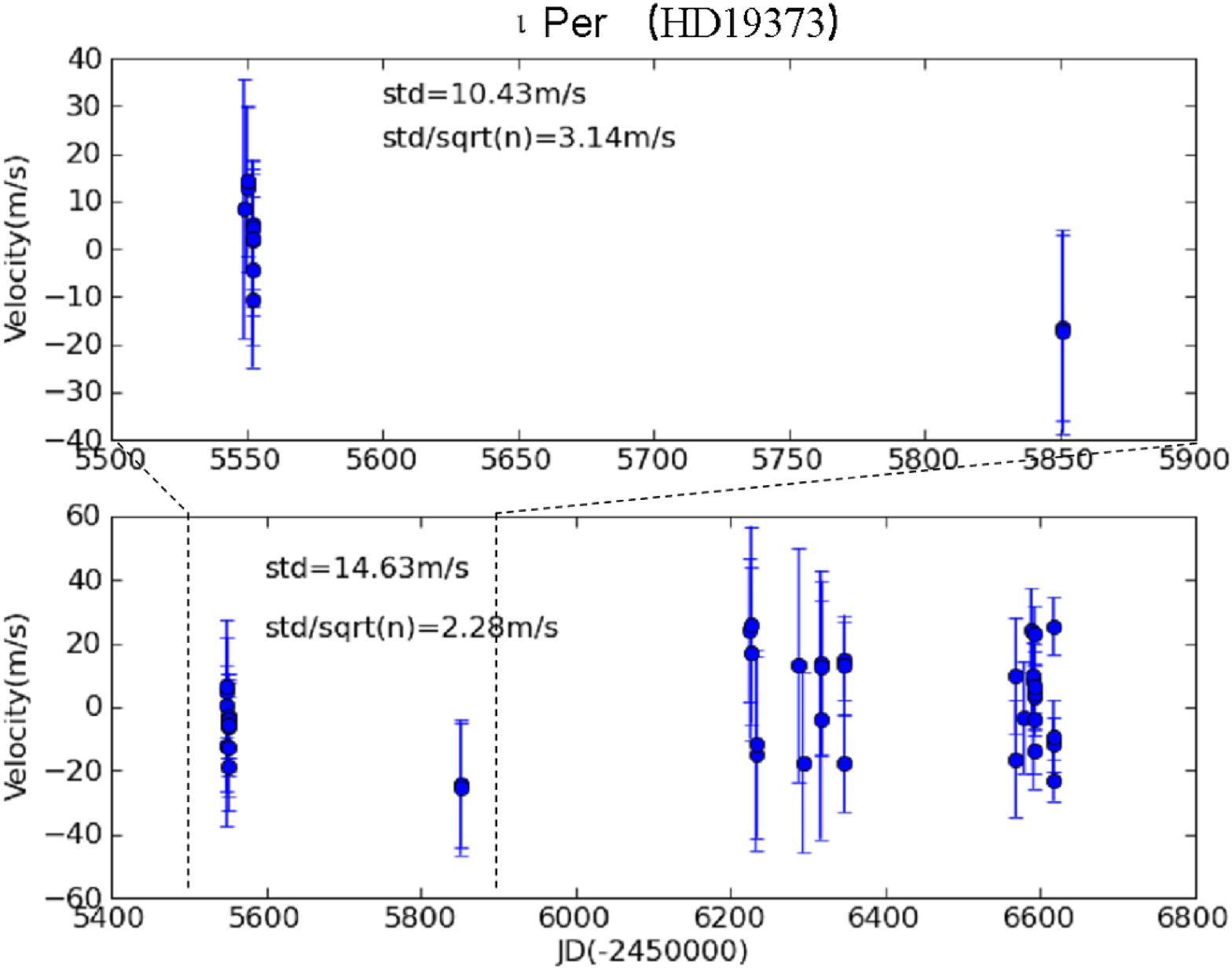}
\caption{The radial velocities of $\iota$ Per (HD19373) obtained from WES. }
\label{FigHD19373}
\end{center}
\end{figure}

We also analyzed the radial velocities of $\iota$ Per (HD\,19373, 
Vmag=4.05, F9.5\uppercase\expandafter{\romannumeral5}) as shown in 
Fig.~\ref{FigHD19373}. The standard deviation of the RVs is 10.43m/s 
over a 10-month period (302 days), with the internal measurement error 
of 3.14 m/s (top panel of Fig.~\ref{FigHD19373}). The standard deviation 
of the RVs is 14.63m/s over a 35-month period (1,068 days), with the 
internal measurement error of 2.28 m/s (bottom panel of 
Fig.~\ref{FigHD19373}).

The measurement errors of RV obtained from each frame are about 2 - 10 
m/s in most of time, but larger (40 - 60 m/s) in JD (-2450000) from 6024 
to 6026 in Fig.~\ref{FigHD113226}, due to the instability of I$_{2}$ 
temperature. The unstability was caused by the problem of conducting 
wire which affected the accuracy of the instrument profile measurement.

Wittenmyer et al. (2015) analyzed 38 observations of $\beta$ Gem (HD 
62509, Vmag=1.14, K0\uppercase\expandafter{\romannumeral3}) over a 
500-day baseline, with a mean internal velocity uncertainty of 8.6 m/s. 
All published velocities spanning more than 25 years are included to fit 
a Keplerian orbit to the planetary signal. The RMS about the fit to the 
velocities obtained from WES is only 7.3 m/s, better than any previous 
published data set, demonstrating that the data acquisition and RV 
extraction techniques of WES are robust.

\subsection{Results about fiber-shaking device}

\begin{figure}
\begin{center}
\includegraphics[angle=0,scale=0.25]{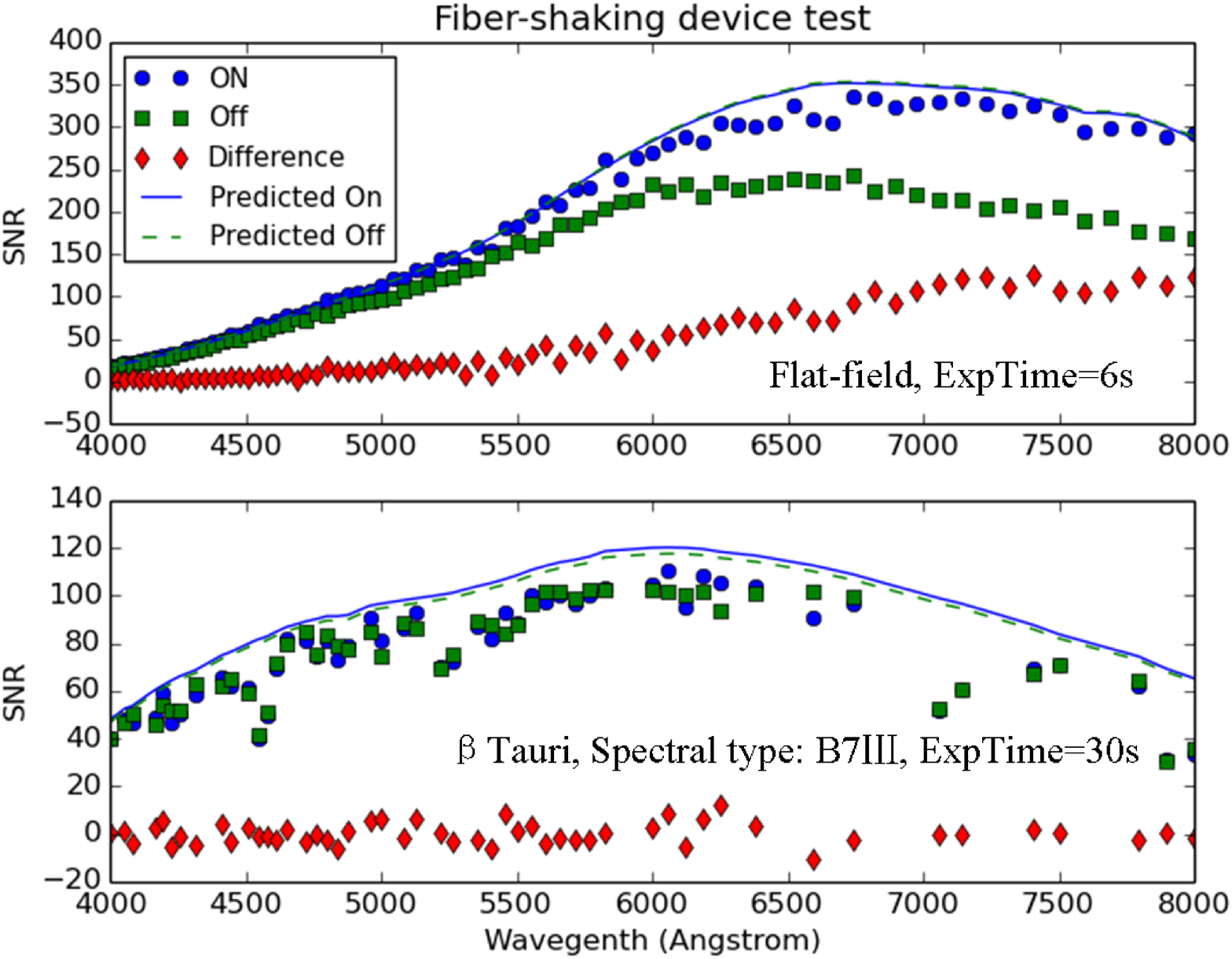}
\caption{Measured and predicted SNR when the fiber-shaking device on and off. Top panel: SNR difference of Flat-field; Bottom panel: SNR difference of B star. }
\label{FigShaker}
\end{center}
\end{figure}

The SNR will be deduced by modal noise which leads to a non-uniform 
distribution of light at the fiber exit. A fiber-shaking device allowing 
non-harmonic movement of the fiber will changes the speckle distribution 
quickly on the fiber exit during exposure time (Sect2.2).  The light 
path of flat-field and ThAr lamp on the fiber-head is stable. The modal 
distribution will not be changed when the fiber-shaking device is off. 
The difference of SNR between the fiber shaking device on and off is 
showen in Fig.~\ref{FigShaker}. The measured average SNR of flat-field 
frame is about 310 and 200 at 7000 - 8000 $\AA$ when the fiber-shaking 
device is on and off respectively (top panel of Fig.~\ref{FigShaker}). 
The measured SNRs almost reach the predicted ones calculated from pure 
photon statistics using the square root of mean photon numbers 
($\sqrt{N}$), expected for a well exposed flat-field spectrum when the 
fiber-shaking device is on.  The fiber shaking device improves the 
scrambling of flat field frames and ThAr frames. The increase of SNR can 
optimize flat fielding and wavelength calibration in data preprocessing.

The light from star at fiber-head changed the input light path at 
high-frequency because of atmospheric perturbation in most seeing 
situations, as the effect of a fiber-shaking device. The modal distribution also 
changed quickly when the fiber-shaking device is off. We did the 
fiber-shaking device test with observations for spectral type B stars 
(see the bottom panel of Fig.~\ref{FigShaker}). The SNR did not change 
significantly when the fiber-shaking device is on and off, and it also 
reached the highest SNR expected with seeing 2$''$. During exposure of 
short time and very good seeing, the SNR may increase or remain 
unchanged when the fiber-shaking device is on, which still need further 
tests.

\section{Summary and future prospects}

\begin{table}
  \begin{center}
  \caption{Main characteristics of WES.}
  \label{TabWESall}
 {
  \begin{tabular}{ll}\hline \hline
\noalign{\smallskip}
{\bf Telescope interface module}		&	\\
\quad Guiding CCD						& SBIG ST2000 XMI, field of view: $\sim6^{\prime}\times6^{\prime}$\\
\quad Pinhole aperture 					& 0.1 mm (2.6$''$) \\ \hline
\noalign{\smallskip}
{\bf Fiber optics} &				\\	
\quad Fiber core & 50 $\mu$m  \\
\quad Entrance focal ratio  &   f/3.67 \\
\quad Exit focal ratio  & $\sim$f/3.7 \\ \hline\noalign{\smallskip}
{\bf Spectrograph} 	&   \scriptsize{white-pupil layout}\\
\quad Collimated beam diameter 				& 92.5 mm\\
\quad Collimator focal ratio				& f/10 \\
\quad Echelle grating 					& Newport R3 (71.5$^{\circ}$), 31.6 grooves/mm, 128 $\times$ 254 mm \\
\quad Cross-disperser						& 2 LF5 prisms, 41$^{\circ}$ apex angle,\\
 								& 103 (base) $\times$ 125 (width) $\times$ 134 (height) mm \\
\quad Dioptric camera						& \\
\quad \quad Camera focal ratio 					& f/3.0 \\
\quad \quad Camera aperture						& 116 mm diameter\\  
\quad Detector							& Andor iKon-L DZ936N-BV, 2K$ \times$ 2K 13.5-$\mu$m pixels \\ 
\quad Total size of spectrograph				& 2.3 (L) $\times$ 1.85 (W) $\times$ 1.5 (H) m \\  \hline\noalign{\smallskip}
{\bf Wavelength range} &\\
   					 		     & 371-976 nm (in continuous 100 orders) \\
								& 976-1,100 nm (in extended 7 orders)\\  
{\bf Resolving power}    		& \scriptsize{at 550nm}\\	
\quad Spectral resolution per pixel   		& 0.044 \AA/pix \\
\quad $R=\lambda/\Delta\lambda$ with slit		& 40,600-57,000 \\\noalign{\smallskip}
\quad  Separation between orders			& 11-26 pixels from red to blue \\\hline \noalign{\smallskip} 		
{\bf Temperature stabilitiy} 			&	\\
\quad Temperature change in one day 			& $\pm$0.03 $^{\circ}$C \\
\quad Temperature change in one week			& $\pm$0.04 $^{\circ}$C \\ 
\noalign{\smallskip} \noalign{\smallskip}
{\bf Limiting magnitudes}		& \scriptsize{at 550nm} \\
\quad The normal mode						& Vmag=8 for SNR=100 in 1-hour exposure 	\\ 
\quad The PRV mode				& Vmag=6 for SNR=110 in 10-min exposure \\ 
\noalign{\smallskip} \noalign{\smallskip}
{\bf Stabilitiy for the PRV mode} 					& \scriptsize{The standard deviation of RV standard star}\\
\quad Short-period						& $\sim$10 m/s in 10 months \\
\quad Long-period						& $<$15 m/s in 54 months (4.4 years)\\ 

\hline
  \end{tabular}
  }
 \end{center}
\end{table}

After commissioning since October 2010, WES has reached the predicted 
performance. The main characteristics of WES are given in 
Table~\ref{TabWESall}. In the normal mode, with a seeing of 1.4$''$, the 
limiting magnitude obtained for 1 hour exposure that produces an SNR of 
more than 100 at 550 nm with the resolution of 40,000 was around Vmag=8. 
In the PRV mode, with the resolution of 50,000 and iodine 
cell, the SNR will reach more than 110 at 550 nm for Vmag=6 in 10-min 
exposure.

The temperature change is 0.04$^\circ$C in one week, much lower than 
expected (0.5$^\circ$C), which is good for PRV observations. The 
limiting factors of RV accuracy seem to be: (1) insufficient SNR; (2) 
unstable optical depth of the I$_{2}$ cell due to unstable temperature 
control; (3) systematic error in the instrument profile (IP), etc. So we 
will focus on the stability of optical element, environment control, and 
guiding telescope carefully in future observations.

Using the RV standard stars ($\epsilon$ Vir, $\iota$ Per) and an 
exoplanet hosting star ($\beta$ Gem), we estimated that RV precision is 
about 10 m/s in short period of time (10 months), showing that WES can 
be used to detect and research giant exoplanets (Cao et al. 2014). The 
standard deviation of RV is better than 15 m/s and the internal 
measurement error is lower than 3 m/s over 4.4 years time. The RV 
precision will be better after optimizing the code for measurement in 
the near future. The WES with RV observing system has joined the East 
Asia exoplanet searching network (EAPSNET) and the Pan-Pacific Planet 
Search (PPPS), collaborating with exoplanet hunters from China, Japan, 
Korea and Australia (Izumiura. 2005; Liu et al. 2009; Sato et al. 2008; 
Wittenmyer et al. 2011), to search for the new worlds around giant and 
dwarf stars.

\acknowledgments
\emph{Acknowledgements.} The WES is supported by the National Natural Science Foundation of China 
and Chinese Academy of Sciences joint fund on astronomy under grant No. 
U1331102, by the National Natural Science Foundation of China under 
grant No. 11333002 and by Sino-German Science Foundation under project 
No. GZ788, GZ1183. This work is partly supported by the Young Scholars Program of Shandong
University, Weihai. We would like to thank Gang Zhao, Xiaojun Jiang and Lei Wang (National Astronomical Observatories, Chinese Academy of 
Sciences) for their 
support to Weihai Observatory. We also thank Bun$'$ei Sato and other 
staff members from Okayama Astrophysical Observatory for their iodine 
cell. We are very grateful to the referee for the useful and constructive comments. 

\begin{appendices}
\section*{Appendix A: ZEMAX optical description }
\setcounter{table}{0}
\renewcommand{\thetable}{A\arabic{table}} 
Table~\ref{TabZEMAX1} summarises the ZEMAX data of all optical
surfaces of the WES. The optical system has
been optimised by considering the measured values of the radii of curvature of the elements.

\begin{table}
\begin{center}
\caption{ZEMAX optical surface data.}
\label{TabZEMAX1}
\scriptsize
\begin{tabular}{rlrrrrrl}
\hline\hline
Surf     &Type      	 &Radius      &Thickness                &Glass      &Diameter          &Conic   &Comment \\ \hline
 OBJ &STANDARD      	 &Infinity         &0                &            &         0       &       0 &  Source \\
   1 &COORDBRK$^a$    &       -         &925              &            &         -       &       - &  Tilt Inbeam \\
STO &STANDARD$^b$     &  -1850          &-925              & MIRROR     & 339.9809        &     -1  & Collimator-1\\
   3 &COORDBRK$^c$    &       -         &    0             &            &         -       &       - &  Littrow twist 1/2 \\
   4 &COORDBRK$^d$    &       -         &    0             &            &         -       &       - &  Blaze tilt\\
   5 &COORDBRK        &       -         &    0             &            &         -       &       - &  90deg tilt of Grating\\
   6 &DGRATING$^e$   	&Infinity          &    0             &  MIRROR    &  393.7426       &       0 &  Echelle Grating\\
   7 &COORDBRK       &       -          &    0             &            &         -       &       - &  90deg backtilt of Grating\\
   8 &COORDBRK       &       -          &    0             &            &         -       &       - &  Blaze backtilt\\
   9 &COORDBRK       &       -          &    0             &            &         -       &       - &  Lttrow twist 2/2\\
  10 &COORDBRK       &       -          &  925             &            &         -       &       - &  Shift back littrow\\
  11 &STANDARD$^b$   &   -1850          & -879             &  MIRROR    &  297.2057       &      -1 &  Collimator-1\\
  12 &STANDARD       &Infinity          &   46             &  MIRROR    &  51.84238       &       0 &  Fold Mirror\\
  13 &STANDARD       &Infinity          &  925             &            &  58.22472       &       0 &  Intermediate Slit\\
  14 &COORDBRK       &       -          &    0             &            &         -       &       - &  \\
  15 &STANDARD$^f$   &   -1850          & -852             &  MIRROR    &  389.1993       &      -1 &  Collimator-2\\
  16 &COORDBRK$^g$   &       -          &    0             &            &         -       &       - &  Center coord\\
  17 &COORDBRK$^h$   &       -          &    0             &            &         -       &       - &  Prism-1\\
  18 &STANDARD       &Infinity          &  -65             &     LF5    &  119.2219       &       0 &  Prism-1 front\\
  19 &COORDBRK$^i$   &       -          &    0             &            &         -       &       - &  Prism-1\\
  20 &STANDARD$^j$   &Infinity          &    0             &            &  156.3835       &       0 &  Prism-1 back\\
  21 &COORDBRK$^h$   &       -         &-79.25             &            &         -       &       - &  Prism-1\\
  22 &COORDBRK$^{h,k}$       &       -         &     0             &            &         -       &       - &  Prism-2\\
  23 &STANDARD       &Infinity         &   -65             &     LF5    &  122.2081       &       0 &  Prism-2 front\\
  24 &COORDBRK$^i$   &       -         &     0             &            &         -       &       - &  Prism-2\\
  25 &STANDARD$^{h,j}$   &Infinity         &     0             &            &   146.435       &       0 &  Prism-2 back\\
  26 &COORDBRK       &       -         &   -60             &            &         -       &       - &  Prism-2\\
  27 &STANDARD       &Infinity         &    -5             &            &       116       &       0 &  \\
  28 &STANDARD       &Infinity         &     0             &            &       116       &       0 &  Circular buffle\\
  29 &STANDARD       & -185.13         & -23.5             & S-FPL53    &       120       & -0.3635 &  Lens 1 front\\
  30 &STANDARD       &   267.9         &-15.16             &            &       120       &       0 &  Lens 1 back\\
  31 &STANDARD       &   231.7         &   -10             &   H-K9L    &       120       &       0 &  Lens 2 front\\
  32 &STANDARD       &-173.868         &   -15             & S-FPL53    &       120       &       0 &  Lens 2 back/Lens 3 front\\
  33 &STANDARD       & 2228.84        &-231.95             &            &       120       &       0 &  Lens 3 back\\
  34 &STANDARD       &    -100        &   -32              & H-QK3L     &      100        &      0  & Lens 4 front\\
  35 &STANDARD       &  301.04        & -30.67             &            &       100       &       0 &  Lens 4 back\\
  36 &STANDARD       &130.3006        &     -8             &    H-F4    &        76       &       0 &  Lens 5 front\\
  37 &STANDARD       &Infinity        & -54.94             &            &        76       &       0 &  Lens 5 back\\
  38 &STANDARD       &Infinity        &      0             &            &  33.59009       &       0 &  \\
  39 &COORDBRK       &       -        &      0             &            &         -       &       - &  \\
  40 &STANDARD       &Infinity        &   -2.3           &LITHOSIL-Q    &        40       &       0 &  Dewar front\\
  41 &STANDARD       &Infinity        &     -9           &    VACUUM    &        40       &       0 &  Dewar back\\
 IMA &STANDARD       &Infinity        &                  &    VACUUM    &  25.65352       &       0 &  CCD\\ \hline
\end{tabular}
\end{center}
\scriptsize{
{\it Notes: $^a$Tilt About X: 7.64; $^b$Y- Decenter: 113.8; $^c$Tilt About X: 0.7; $^d$Tilt About Y: 71.5; $^e$X- Decenter: 123.76; $^f$Y- Decenter: -145.6; $^g$Decenter Y: -124; Tilt About X: 1.4; $^h$Tilt About X: -33.8; $^i$Tilt About X: 41; $^j$Y- Decenter: 24.256232; $^k$Decenter Y: 20.6; }}
\end{table}  

\end{appendices}

\end{document}